\documentclass[preprint,showpacs,preprintnumbers,superscriptaddress,amsmath,amssymb]{revtex4}
\usepackage{graphicx}
\usepackage{dcolumn}
\usepackage{bm}
\usepackage[american]{babel}
\begin{document}

\title{Limits on Low Energy Photon-Photon Scattering from an Experiment on Magnetic Vacuum Birefringence}

\author{M.~Bregant}
\affiliation{INFN, sezione di Trieste and Dipartimento di Fisica, Universit\`a di Trieste, Via Valerio 2, 34127 Trieste, Italy}
\author{G.~Cantatore}
\affiliation{INFN, sezione di Trieste and Dipartimento di Fisica, Universit\`a di Trieste, Via Valerio 2, 34127 Trieste, Italy}
\author{S.~Carusotto}
\affiliation{INFN, Sezione di Pisa and Dipartimento di Fisica, Universit\`a di Pisa, Via Buonarroti 2, 56100 Pisa Italy}
\author{R.~Cimino}
\affiliation{INFN, Laboratori Nazionali di Frascati, Via E. Fermi 40, 00044 Frascati, Italy}
\author{F.~Della Valle}
\affiliation{INFN, sezione di Trieste and Dipartimento di Fisica, Universit\`a di Trieste, Via Valerio 2, 34127 Trieste, Italy}
\author{G.~Di Domenico}
\affiliation{INFN, Sezione di Ferrara and Dipartimento di Fisica, Universit\`a di Ferrara, Polo Scientifico, Via Saragat 1 C, 44100 Ferrara, Italy}
\author{U.~Gastaldi}
\affiliation{INFN, Laboratori Nazionali di Legnaro, viale dell'Universit\`a 2, 35020 Legnaro}
\author{M.~Karuza}
\affiliation{INFN, sezione di Trieste, Via Valerio 2, 34127 Trieste, Italy}
\author{V.~Lozza}
\affiliation{INFN, sezione di Trieste and Dipartimento di Fisica, Universit\`a di Trieste, Via Valerio 2, 34127 Trieste, Italy}
\author{E.~Milotti}
\affiliation{INFN, sezione di Trieste and Dipartimento di Fisica, Universit\`a di Trieste, Via Valerio 2, 34127 Trieste, Italy}
\author{E.~Polacco}
\affiliation{INFN, Sezione di Pisa and Dipartimento di Fisica, Universit\`a di Pisa, Via Buonarroti 2, 56100 Pisa Italy}
\author{G.~Raiteri}
\affiliation{INFN, sezione di Trieste and Dipartimento di Fisica, Universit\`a di Trieste, Via Valerio 2, 34127 Trieste, Italy}
\author{G.~Ruoso}
\affiliation{INFN, Laboratori Nazionali di Legnaro, viale dell'Universit\`a 2, 35020 Legnaro}
\author{E.~Zavattini\footnote{Deceased January 9, 2007}}
\affiliation{INFN, sezione di Trieste and Dipartimento di Fisica, Universit\`a di Trieste, Via Valerio 2, 34127 Trieste, Italy}
\author{G.~Zavattini}
\affiliation{INFN, Sezione di Ferrara and Dipartimento di Fisica, Universit\`a di Ferrara, Polo Scientifico, Via Saragat 1 C, 44100 Ferrara, Italy}
\collaboration{PVLAS Collaboration}
\noaffiliation

\date{\today}

\begin{abstract}
Experimental bounds on induced vacuum magnetic birefringence can be used to improve present photon-photon scattering limits in the electronvolt energy range. Measurements with the PVLAS apparatus (E. Zavattini {\it et al.}, Phys. Rev. D {\bf77} (2008) 032006) at both $\lambda = 1064$~nm and $532$~nm lead to bounds on the parameter {\it A$_{e}$}, describing non linear effects in QED, of $A_{e}^{(1064)} < 6.6\cdot10^{-21}$~T$^{-2}$~@~$1064$~nm and $A_{e}^{(532)} < 6.3\cdot10^{-21}$~T$^{-2}$~@~$532$~nm, respectively, at 95\% confidence level, compared to the predicted value of $A_{e}=1.32\cdot10^{-24}$~T$^{-2}$. The total photon-photon scattering cross section may also be expressed in terms of $A_e$, setting bounds for unpolarized light of $\sigma_{\gamma\gamma}^{(1064)} < 4.6\cdot10^{-62}$~m$^{2}$ and $\sigma_{\gamma\gamma}^{(532)} < 2.7\cdot10^{-60}$~m$^{2}$. Compared to the expected QED scattering cross section these results are a factor of $\simeq2\cdot10^{7}$ higher and represent an improvement of a factor about 500 on previous bounds based on ellipticity measurements and of a factor of about $10^{10}$ on bounds based on direct stimulated scattering measurements.
\end{abstract}

\pacs{12.20.Fv, 07.60.Fs, 14.80.Mz}

\maketitle

\section{Introduction}
Classical electrodynamics in vacuum is a linear theory and does not foresee photon-photon scattering or other non linear effects between electromagnetic fields. Before quantum electrodynamics (QED) was formally complete, Euler and Heisenberg, and Weisskopf, realized that vacuum fluctuations, permitted by the uncertainty principle, lead to non linear effects: 4 photons can couple via fermion loops. Such non linear effects were first calculated in 1936 \cite{QED} and can be described by an effective Lagrangian $L_{EHW}$ which, for field strengths well below their critical values ($B \ll B_{\rm crit}={m^{2}c^{2}}/{e \hbar}=4.4\cdot10^{9}$~T, $E \ll E_{\rm crit}={m^{2}c^{3}}/{e \hbar}=1.3\cdot10^{18}$~V/m) and for photon energies below the electron mass, can be written as (in S.I. units):
\begin{eqnarray}
L_{\rm EHW}& =& \frac{A_{e}}{\mu_{0}}\bigg[\left(\frac{E^2}{c^2}-B^2\right)^2+7\left(\frac{\vec{E}}{c}\cdot\vec{B}\right)^2\bigg]
\end{eqnarray}
where the parameter $A_e$ describing the non linearity is
\begin{equation}
A_e=\frac{2}{45\mu_{0}}\frac{\alpha^2 \mathchar'26\mkern-10mu\lambda_e^{3}}{m_{e}c^{2}}=1.32\cdot10^{-24} {\text{~T}}^{-2}
\end{equation}
with $\mathchar'26\mkern-10mu\lambda_e$ being the Compton wavelength of the electron, $\alpha={e^2}/{(\hbar c 4\pi\epsilon_0)}$ the fine structure constant, $m_e$ the electron mass, $c$ the speed of light in vacuum and $\mu_0$ the magnetic permeability of vacuum.

Maxwell's equations are still valid provided the constitutive equations are applied to the total Lagrangian density $L=L_{\rm Class}+L_{\rm EHW}$ to derive the displacement vector $\vec{D}$ and the magnetic field intensity vector $\vec{H}$
\begin{eqnarray}
\vec{D}&=&\frac{1}{\epsilon_{0}}\frac{\partial L}{\partial\vec{E}}\nonumber\\
\vec{H}&=&-{\mu_{0}}\frac{\partial L}{\partial\vec{B}}
\label{constitutive}
\end{eqnarray}

\begin{figure}[htb]
\begin{center}
{\includegraphics*[width=7cm]{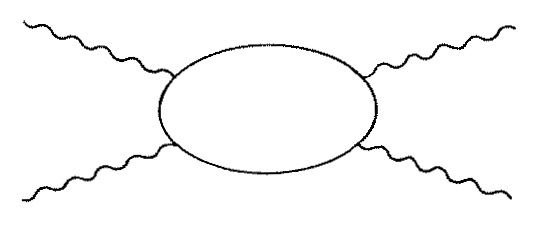}
\includegraphics*[width=7cm]{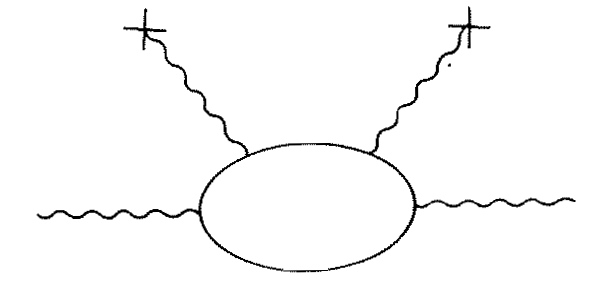}}
\caption{\it First order Feynman diagrams for both photon-photon scattering and magnetically (electrically) induced vacuum birefringence.}
\label{diag}
\end{center}
\vspace{-0.6cm}
\end{figure}

One of the yet to be measured effects predicted by the $L_{EHW}$ correction is that vacuum will become birefringent in the presence of an external magnetic and/or electric field. For example, in the case of a beam propagating perpendicularly to an external magnetic field, if $n_{\parallel}$ and $n_{\perp}$ indicate the index of refraction for 
polarizations respectively parallel and perpendicular to the field direction, the birefringence can be expressed as \cite{Adler,Iacopini}
\begin{equation}
n_{\parallel}-n_{\perp}=\Delta n^{\rm (QED)} = 3 A_e B^2
\label{birifqed}
\end{equation}
which is extremely small: with a field intensity of 5~T, $\Delta n^{\rm (QED)}\approx10^{-22}$.
Another process described by the same Feynman diagrams as magnetically induced vacuum birefringence is photon-photon scattering. Figure \ref{diag} shows the Feynman diagrams for both photon-photon scattering and field induced vacuum birefringence. 
In general, the connection between the index of refraction $n$ of a medium and the photon scattering amplitude in the forward direction for photons with energy $E_{\gamma}$, $f(\vartheta=0,E_\gamma)$, is (see for example \cite{Newton})
\begin{equation}
n=1+\frac{2\pi}{k^2}Nf(0,E_{\gamma})
\end{equation}
where $N$ is the average number density of centers of scattering and $k$ is the photon wave number. Applied to photon-photon scattering
of linearly polarized photons, the center of mass forward scattering amplitude of ingoing and outgoing photons all having parallel polarizations, $f^{\rm (QED)}_{\parallel}(0,E_{\gamma})$, and the one in which the two incoming photons have perpendicular polarizations as do the ougoing photons, $f^{\rm (QED)}_{\perp}(0,E_{\gamma})$ are, respectively \cite{Haissinski}
\begin{eqnarray}
f^{\rm (QED)}_{\parallel}(0,E_{\gamma})&=&\frac{32}{45}\frac{\alpha^2\mathchar'26\mkern-10mu\lambda_e}{4\pi}\left(\frac{E_{\gamma}}{m_e c^2}\right)^{3} = \frac{16\mu_{0}}{4\pi\hbar^{2}c^{2}}A_e {E_{\gamma}}^3\\
f^{\rm (QED)}_{\perp}(0,E_{\gamma})&=&\frac{56}{45}\frac{\alpha^2\mathchar'26\mkern-10mu\lambda_e}{4\pi}\left(\frac{E_{\gamma}}{m_e c^2}\right)^{3}= \frac{28\mu_{0}}{4\pi\hbar^{2}c^{2}}A_e {E_{\gamma}}^3
\end{eqnarray}
where it is apparent that the scattering amplitude is proportional to $A_e$. The authors of \cite{Haissinski} also show that $N$ is proportional to the energy density of the scatterer field (electric and/or magnetic) and inversely proportional to the photon energy in the center of mass reference frame. From the scattering amplitude one can find the differential cross section
\begin{equation}
\frac{d\sigma_{\gamma\gamma}}{d\Omega}(\vartheta,E_{\gamma})=|f(\vartheta,E_{\gamma})|^{2}
\end{equation}
and the total cross section which depends on $A_{e}^2$. For unpolarized light one finds \cite{DeTollis,Duane,Karplus,Bernard,BernardOld}
\begin{equation}
\sigma_{\gamma\gamma}^{\rm (QED)}(E_{\gamma})=\frac{1}{45^2}\frac{973}{5\pi}\alpha^4\left(\frac{E_{\gamma}}{m_{e}c^{2}}\right)^{6}{\mathchar'26\mkern-10mu\lambda_e^{2}}=\frac{973\mu_{0}^{2}}{20\pi}\frac{E_{\gamma}^{6}}{\hbar^{4}c^{4}}A_{e}^{2}
\end{equation}
The connection between the total photon-photon cross section and vacuum birefringence, hence the parameter $A_e$ describing non linear QED effects, makes non linear QED searches via ellipsometric techniques very attractive. Limits on $A_e$ from ellipsometric data can therefore be directly translated into photon-photon scattering limits.

It is interesting to note that in a post-Maxwellian framework \cite{Denisov} the Lagrangian density $L_{\rm pM}$ describing nonlinear electrodynamic effects in vacuum is parameterized by three parameters $\xi$, $\eta_{1}$ and $\eta_{2}$:
\begin{eqnarray}
L_{\rm pM}& =& \frac{\xi}{2\mu_{0}}\bigg[\eta_{1}\left(\frac{E^2}{c^2}-B^2\right)^2+4\eta_{2}\left(\frac{\vec{E}}{c}\cdot\vec{B}\right)^2\bigg]
\label{Lpm}
\end{eqnarray}
In this parameterization $\xi=1/B_{\rm crit}^{2}$, and $\eta_{1}$ and $\eta_{2}$ are dimensionless parameters depending on the chosen model. 
In the Euler-Heisenberg electrodynamics $\eta_{2}^{\rm (QED)}=\frac{7}{4}\eta_{1}^{\rm (QED)}=\alpha/(45\pi)$, $\alpha$ being the fine structure constant. 

By substituting the post-Maxwellian generalization into equations (\ref{constitutive}) one finds that the birefringence induced by a transverse magnetic field is (to be compared with equation (\ref{birifqed}))
\begin{equation}
\Delta n^{\rm (pM)} = 2\xi(\eta_{2}-\eta_{1})B^{2}
\end{equation}
whereas the forward scattering amplitudes given in expressions (6) and (7) become
\begin{eqnarray}
f^{\rm (pM)}_{\parallel}(0,E_{\gamma})&=& \frac{8\mu_{0}}{4\pi\hbar^{2}c^{2}}\xi\eta_{1} {E_{\gamma}}^3\\
f^{\rm (pM)}_{\perp}(0,E_{\gamma})&=& \frac{8\mu_{0}}{4\pi\hbar^{2}c^{2}}\xi\eta_{2} {E_{\gamma}}^3
\end{eqnarray}
Birefringence is therefore only sensitive to the difference $\eta_{2}-\eta_{1}$ whereas the two forward scattering amplitudes $f^{\rm (pM)}_{\parallel}(0,E_{\gamma})$ and $f^{\rm (pM)}_{\perp}(0,E_{\gamma})$ are proportional respectively to $\eta_{1}$ and $\eta_{2}$. At scattering angles different from $\vartheta=0$ it remains true that $f^{\rm (pM)}_{\parallel}(\vartheta,E_{\gamma})$ is proportional to $\eta_{1}$ but $f^{\rm (pM)}_{\perp}(\vartheta,E_{\gamma})$ will now depend on a combination of $\eta_{1}$ and $\eta_{2}$ which never cancels.
Therefore, for example, in the Born-Infeld model \cite{Born} where $\eta_{1}=\eta_{2}$, magnetically induced birefringence is not expected even though photon-photon scattering is.
Although very promising for detecting nonlinear electrodynamic effects, the ellipsometric technique alone it is not sufficient to determine the two independent quantities $\xi\eta_{1}$ and $\xi\eta_{2}$. On the other hand direct photon-photon scattering with defined polarization states can. It is clear how both techniques are complementary.

Assuming the Euler-Heisenberg Lagrangian density, in this paper we will present the best limits on $\sigma_{\gamma\gamma}$ at low energy available today.

\section{Apparatus and Method}
\begin{figure}[htb]
\begin{center}
{\includegraphics*[width=14cm]{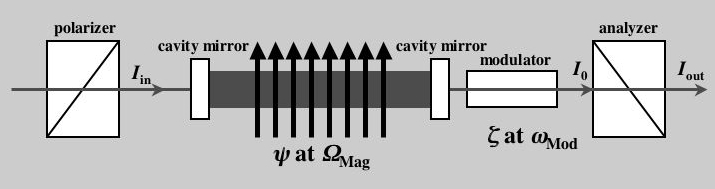}}
\caption{\it Schematic layout of a sensitive ellipsometer. See text for description.}
\label{SchemePVLAS}
\end{center}
\vspace{-0.6cm}
\end{figure}

The general scheme of a sensitive ellipsometer searching for magnetically induced birefringence is presented in Figure \ref{SchemePVLAS}. 
A polarizer defines the polarization of the beam, of power $I_{\rm in}$, before it enters the magnetic field region where it acquires an ellipticity $\psi$. The ellipticity is
made time dependent by modulating the magnetic field with angular frequency
$\Omega_{\rm Mag}$ (see text below). Two mirrors compose either a multi-pass or a Fabry-Perot cavity to increase the optical path within the magnetic field region. The beam then passes first through a modulator, where it acquires a known ellipticity $\zeta$ modulated at frequency $\omega_{\rm Mod}$, and then through an analyzer. The transmitted power $I_{\rm out}$ is then detected and analysed.

\subsection{Heterodyne technique}
\begin{figure}
\begin{center}
{\includegraphics*[width=10cm]{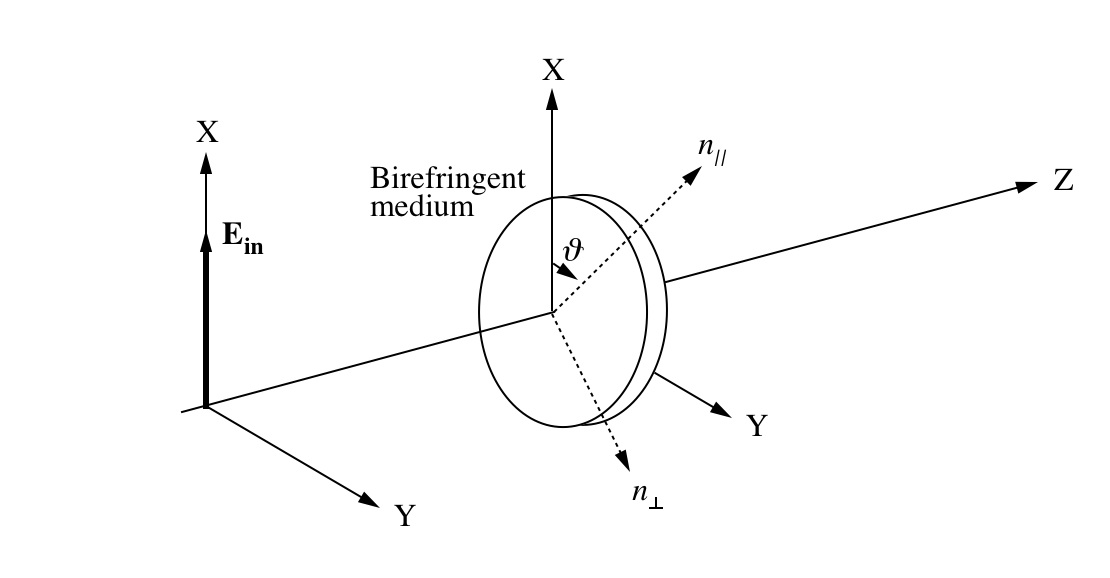}}
\caption{\it Reference frame for the calculations using the Jones matrix formalism.}
\label{SistRif}
\end{center}
\end{figure}

For the purpose of our discussion let a laser beam propagate along the $Z$ axis and let the incoming (linear) polarization define the $X$ axis (Figure \ref{SistRif}). Considering the coherence of our light source, the Jones matrix formalism will be used. The Jones matrix for a uniaxial birefringent element is given by
\begin{equation}
\mathbf{BF}(\vartheta)=
\left(
\begin{array}{cc}
1+\imath\psi\cos{2\vartheta} & \imath\psi\sin{2\vartheta} \\
 \imath\psi\sin{2\vartheta} & 1-\imath\psi\cos{2\vartheta}
 \end{array}
 \right)
 \end{equation}
where $\vartheta$ represents the angle between the slow axis ($n_\parallel > n_{\perp}$) of the medium and the $X$ axis and $\psi\ll1$ is the induced ellipticity acquired by the light, given by 
\begin{equation}
\psi=\frac{\varphi_{\parallel}-\varphi_{\perp}}{2}=\pi\frac{ L(n_{\parallel}-n_{\perp})}{\lambda}
\end{equation}
with $\varphi_{\parallel}-\varphi_{\perp}$ the phase delay between the parallel and perpendicular polarization components acquired in a length $L$.

Given an input beam whose electric field after the entrance polarizer is $\vec{E}_{\rm in}=E_{\rm in}{1 \choose 0}$ the electric field after the birefringent medium will be
\begin{eqnarray}
\nonumber
\vec{E}_{\rm 0}&=&E_{\rm in}\mathbf{\cdot BF}\cdot{1 \choose 0} = E_{\rm in}
\left(
\begin{array}{c}
1+\imath\psi\cos{2\vartheta} \\
\imath\psi\sin{2\vartheta} 
\end{array}
\right)
\end{eqnarray}
Assuming no losses, the power $I_{\rm out}$ after the analyzer (polarizer crossed with respect to the entrance polarizer) will therefore be
\begin{equation}
I_{\rm out}=I_{\rm in}\left|\imath\psi\sin{2\vartheta}\right|^{2}
\end{equation}

The power is proportional to $\psi^2$ and, whether $\vartheta$ is constant in time or not, results in an unmeasurably small intensity component.

To linearize the term proportional to the ellipticity signal $\psi$ to be detected, one can add a known time varying ellipticity $\zeta(t)$ using an ellipticity modulator. The Jones matrix for the modulator is the same as $\mathbf{BF}$ set at an angle of $\pi/4$ ($\psi\ll\zeta\ll1$):
\begin{equation}
\mathbf{MOD}=
\left(
\begin{array}{cc}
1& \imath\zeta(t) \\
 \imath\zeta(t) & 1
 \end{array}
 \right)
 \end{equation}
and the resulting vector describing the electric field after the modulator will be 
\begin{eqnarray}
&\vec{E}_{\rm 0}=E_{\rm in}\mathbf{\cdot MOD \cdot BF}\cdot{1 \choose 0}\\
\nonumber\\
=&E_{\rm in}
\left(
\begin{array}{c}
1+\imath\psi\cos{2\vartheta}-\psi\zeta(t)\sin2\vartheta \\
\imath\zeta(t)+\imath\psi\sin{2\vartheta}-\zeta(t)\psi\cos2\vartheta
\end{array}
\right)
\end{eqnarray}
Neglecting second order terms, the power $I_{\rm out}$ after the analyzer will be
\begin{equation}
I_{\rm out}(t)=I_{\rm in}\left|\imath\zeta(t)+\imath\psi\sin{2\vartheta}\right|^{2} \simeq I_{\rm in}\left[\zeta(t)^2+2\zeta(t)\psi\sin2\vartheta\right]
\end{equation}
which now depends linearly on the ellipticity $\psi$. To complete the discussion, one finds experimentally that static and slowly varying ellipticities, in the following indicated as $\alpha(t)$, are always present in an actual apparatus and that two crossed polarizers have an intrinsic extinction ratio $\sigma^2$, mainly due to imperfections in the crystals they are made of. Furthermore, losses in the system reduce the total light reaching the analyzer. Therefore, taking into account an additional 
spurious ellipticity term $\alpha(t)$ (since $\alpha, \psi, \zeta \ll 1$ these terms commute and therefore add up algebraically) and a term proportional to $\sigma^2$, the total power at the output of the analyzer will be
\begin{eqnarray}
\nonumber
I_{\rm out}(t)&=&I_{0}\left[\sigma^2+\left|\imath\zeta(t)+\imath\psi\sin{2\vartheta}+\imath\alpha(t)\right|^{2}\right]\simeq\\
&\simeq&I_{0}\left[\sigma^2+\zeta(t)^2+\alpha(t)^{2}+2\zeta(t)\psi\sin2\vartheta+2\zeta(t)\alpha(t)\right]
\end{eqnarray}
where $I_{0}$ represents the power of light reaching the analyser.

To be able to distinguish the term $\zeta(t)\alpha(t)$, which is usually largest at low frequencies, from the term $\zeta(t)\psi\sin2\vartheta$, the term of interest $\psi\sin2\vartheta$ is also made to vary in time. This can be done by either ramping the magnetic field intensity (varying therefore $\psi$) or by rotating the magnetic field direction (varying $\vartheta$). The final expression, explicitly indicating the time dependence of $\psi$ and $\vartheta$, for the power at the output of the analyzer is therefore
\begin{equation}
I_{\rm out}(t)=I_{0}\left[\sigma^2+\zeta(t)^2+\alpha(t)^2+2\zeta(t)\psi(t)\sin2\vartheta(t)+2\zeta(t)\alpha(t)\right]
\label{spurious}
\end{equation}
\subsection{Optical path multiplier}
To further increase the ellipticity induced by the birefringent region one can increase the number of passes through it. Either a multi-pass cavity or a Fabry-Perot cavity can be used for this purpose. In the PVLAS experiment described below, a Fabry-Perot has been chosen.
In a multi-pass cavity the induced ellipticity is proportional to the number of passes $N_{\rm pass}$ through the region. With a Fabry-Perot cavity the calculation is
 not immediate since one is dealing with a standing wave.

Let $t$, $r$ be the transmission and reflection coefficients, and $p$ the losses of the mirrors of the cavity such that $t^2+r^2+p=1$. Let $d$ be the length of the cavity and $\delta={4\pi d}/{\lambda}$ the roundtrip phase for a beam of wavelength $\lambda$. Then the Jones matrix for the elements of the ellipsometer after the entrance polarizer is
\begin{equation}
\mathbf{ELL}=
\mathbf{A}\cdot\mathbf{SP}\cdot\mathbf{MOD}\cdot{\it t}^{2}e^{\imath\delta/2}\sum_{n=0}^{\infty} {\left[\mathbf{BF}^{2}{\it r}^{2}e^{\imath\delta}\right]^{n}}\cdot\mathbf{BF}
 \end{equation} 
where $\mathbf{A}=\left(
\begin{array}{cc}
0 & 0 \\
0 & 1
 \end{array}
 \right)$ is the analyzer Jones matrix and $\mathbf{SP}$ describes the spurious ellipticity.
 Because ${\it r}^{2} < 1$, $\mathbf{ELL}$ can be rewritten as
\begin{equation}
\mathbf{ELL}=
\mathbf{A}\cdot\mathbf{SP}\cdot\mathbf{MOD}\cdot{\it t}^{2}e^{\imath\delta/2}{\left[\mathbf{I}-\mathbf{BF}^{2}{\it r}^{2}e^{\imath\delta}\right]^{-1}}\cdot\mathbf{BF}
\label{fp}
\end{equation} 
with $\mathbf{I}$ the identity matrix. With the laser phase locked to the cavity so that $\delta=2\pi m$, where $m$ is an integer number, 
the electric field at the output of the system will be
\begin{equation}
\vec{E}_{\rm out}=E_{\rm in}\mathbf{\cdot ELL}\cdot{1 \choose 0}
=E_{\rm in}\frac{{\it t}^{2}}{{\it t}^{2}+p}
\left(
\begin{array}{c}
0 \\
\imath\alpha(t)+\imath\zeta(t)+\imath\frac{1+{\it r}^{2}}{1-{\it r}^{2}}\psi\sin{2\vartheta}
\end{array}
\right)
\end{equation}
and the power, including losses,
\begin{equation}
I_{\rm out}(t)=I_{0}\Bigg|\imath\alpha(t)+\imath\zeta(t)+\imath\left(\frac{1+{\it r}^{2}}{1-{\it r}^{2}}\right)\psi\sin{2\vartheta}\Bigg|^{2}
\label{Iout}
\end{equation}
This expression is at the basis of the ellipsometer in the PVLAS apparatus.
Small ellipticities add up algebraically and the Fabry-Perot multiplies the single pass ellipticity $\psi\sin{2\vartheta}$, generated within the cavity, by a factor $({1+{\it r}^{2}})/({1-{\it r}^{2}})\approx{2{\cal F}}/{\pi}$, where ${\cal F}$ is the finesse of the cavity. The ellipticity signal to be detected is therefore $\Psi = ({2{\cal F}}/{\pi})\psi\sin{2\vartheta}$. Typical values for the finesse ${\cal F}$ of the PVLAS cavity are $\simeq10^5$.

In the PVLAS experiment, $\zeta(t)=\zeta_{0}\cos(\omega_{\rm Mod}t+\theta_{\rm Mod})$ and the magnetic field direction is rotated at an angular velocity $\Omega_{\rm Mag}$. A Fourier analysis of the power $I_{\rm out}(t)$ of equation (\ref{Iout}) results in four main frequency components each with a definite amplitude and phase. These are reported in table \ref{components}.
\begin{table}[h!]
\caption{Intensity of the frequency components of the signal after the analyzer $\mathbf{A}$.}
\begin{tabular}{c|c|c|c}
\hline\noalign{\smallskip}
Frequency & Fourier component & Intensity/$I_{0}$ & Phase\\
\noalign{\smallskip}\hline\noalign{\smallskip}
$DC$ & $I_{\rm DC}$ & $\sigma^2+\alpha_{\rm DC}^2+\zeta_{0}^{2}/2$ & $-$\\
$\omega_{\rm Mod}$ & $I_{\omega_{\rm Mod}}$ & $2\alpha_{\rm DC}\zeta_{0}$ & $\theta_{\rm Mod}$\\
$\omega_{\rm Mod}\pm2\Omega_{\rm Mag}$ & $I_{\omega_{\rm Mod}\pm2\Omega_{\rm Mag}}$ & $\zeta_{0}\frac{2{\cal F}}{\pi}\psi$ & $\theta_{\rm Mod}\pm2\theta_{\rm Mag}$\\
$2\omega_{\rm Mod}$ & $I_{2\omega_{\rm Mod}}$ & $\zeta_{0}^{2}/2$ & $2\theta_{\rm Mod}$\\
\noalign{\smallskip}\hline
\end{tabular}
\label{components}
\end{table}

The presence of a component at $\omega_{\rm Mod}\pm2\Omega_{\rm Mag}$ in the signal identifies an induced ellipticity within the Fabry-Perot cavity. Furthermore the phase of this component must satisfy the value in table \ref{components}.

\subsection{PVLAS Apparatus}
\begin{figure}[htb]
\begin{center}
{\includegraphics*[width=10cm]{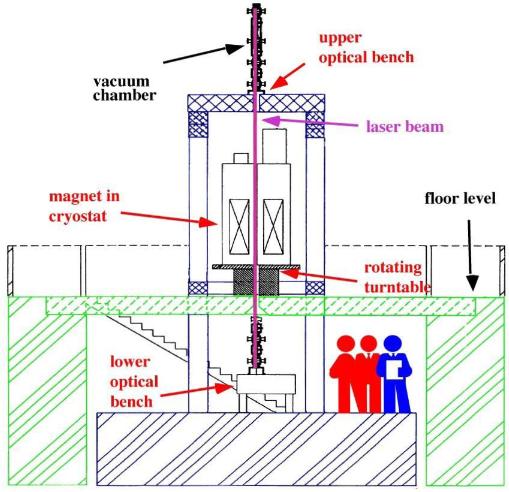}}
\caption{\it Schematic layout of the PVLAS apparatus. See text for description.}
\label{Apparato}
\end{center}
\vspace{-0.6cm}
\end{figure}

A description of the PVLAS apparatus, shown schematically in Figure \ref{Apparato}, can be found in \cite{PRD, PRL, HypIn}. The magnetic field is provided by a superconducting dipole magnet which is placed vertically and rotates around its axis, at a typical frequency of 0.3 Hz. The magnetic field therefore lies in the horizontal plane, the field region is 1 m long, and the maximum field intensity is 5.5 T.
The ellipsometer develops vertically: polarizer and entrance cavity mirror are supported by the lower optical bench, whereas  output cavity mirror,  modulator and analyzer are in a vacuum chamber on the upper optical bench. The lower optical bench is in a pit whose floor is a concrete slab resting on four 14 m long pillars buried in the ground. The slab and pillars are therefore seismically isolated with respect to the surrounding hall floor and building. The upper optical bench is sustained by a granite tower 7 m high also standing on the concrete slab. The upper and lower vacuum chambers are connected by a quartz tube 2.5 cm in diameter which passes through the warm bore of the cryostat containing the magnet. The magnet and turntable are supported by a concrete beam crossing over the pit and resting on the hall floor. Thus, mechanical vibrations due to the rotating magnet reaching the optical system will be greatly suppressed and should not cause excess ellipticity noise. 

The vacuum system is based on two liquid N$_2$ traps combined with Ti sublimation getter, and pressure is kept at the level of $P \approx 10^{-8}$ mbar during measurements. For test purposes the vacuum chamber can be filled with gases at known pressure,  measured with a set of capacitive transducers. The presence of the gas gives rise to a known magnetic birefringence via the Cotton Mouton effect \cite{rizzo,O2,Xe}.

The laser source is frequency locked to the Fabry-Perot cavity using a modified Pound-Drever-Hall technique \cite{rsi}. Two different light sources were  alternatively used: an infrared Nd:YAG laser emitting 800 mW at 1064~nm (infrared),  and its frequency doubled secondary output of 80 mW at 532~nm (green). The cavity parameters were as follows: finesse ${\cal F}_{1064}=70000$, output power $P_{1064}$= 60 mW for the infrared and ${\cal F}_{532}=37000$, $P_{532}$=1.5 mW for the green. 

The light transmitted by the analyzer is detected by a photodiode connected to a low noise current amplifier and the signal is then sent to both a spectrum analyzer, for online monitoring of the apparatus, and to a lock-in amplifier demodulated at $\omega_{\rm Mod}$. To make the analysis independent from the instability of the rotation frequency of the turntable sustaining the magnet, the table perimeter is equipped with 32 equally spaced trigger marks. The output of the lock-in amplifier is acquired at the passage of each trigger mark, therefore maintaining the coherence of the searched signal even for long integration times.

\subsection{Noise considerations}
In the presence of a signal above background with the correct Fourier phase, the ellipticity $\Psi= ({2{\cal F}}/{\pi})\psi$ can be calculated from $I_{0}$, from the Fourier components $I_{\omega_{\rm Mod}\pm2\Omega_{\rm Mag}}$, and from $I_{2\omega_{\rm Mod}}$  as the average of the two sideband signals:

\begin{equation}
\Psi=
\frac{1}{2}\left(\frac{I_{\omega_{\rm Mod}+2\Omega_{\rm Mag}}}{\sqrt{2I_{0}I_{2\omega_{\rm Mod}}}}+\frac{I_{\omega_{\rm Mod}-2\Omega_{\rm Mag}}}{\sqrt{2I_{0}I_{2\omega_{\rm Mod}}}}\right)
\end{equation}

Indicating with $R_{\omega_{\rm Mod}\pm2\Omega_{\rm Mag}}$ the noise spectral density at the signal frequencies, and assuming $R_{\omega_{\rm Mod}+2\Omega_{\rm Mag}}=R_{\omega_{\rm Mod}-2\Omega_{\rm Mag}}$, the sensitivity spectral density $\Psi_{\rm Sens}$ of the ellipsometer for a unity signal to noise ratio is
\begin{equation}
\Psi_{\rm Sens}=
\frac{R_{\omega_{\rm Mod}+2\Omega_{\rm Mag}}}{\sqrt{4I_{0}I_{2\omega_{\rm Mod}}}}
\label{sensit}
\end{equation}

In principle the r.m.s. noise limit for such a system is determined by the r.m.s. shot-noise $i_{\rm shot}$ of the DC current $i_{\rm DC}$ generated by the modulation amplitude $I_{0}q\zeta_{0}^{2}/{2}$,
 by the extinction ratio $I_0q\sigma^2$ and the by DC component of the spurious ellipticity $I_0q\alpha^2_{\rm DC}$ (see table \ref{components}):
\begin{equation}
i_{\rm shot}=\sqrt{{2ei_{\rm DC}\Delta \nu}}=\sqrt{{2eI_{0}q(\sigma^2+\frac{{\zeta_0}^2}{2}+\alpha^2_{\rm DC})\Delta \nu}}
\end{equation}
where $q$ is the quantum efficiency of the photodetector, $\Delta \nu$ is the bandwidth and $e$ is the electron charge. In the case $\zeta_{0}^2 \gg \sigma^2$ and  $\zeta_{0}^2 \gg \alpha_{\rm DC}^2$ the DC current will only depend on $\zeta_0$
and by substituting $R_{\omega_{\rm Mod}\pm\Omega_{\rm Mag}} = i_{\rm shot}/(q\sqrt{\Delta \nu})$ into equation (\ref{sensit}) the shot-noise sensitivity spectral density $\Psi_{\rm shot}$ becomes
\begin{equation}
\Psi_{\rm shot}=\sqrt{\frac{e}{2I_{0}q}}
\end{equation}

For a power $I_{0}=10$ mW and a quantum efficiency $q = 0.7$~A/W this leads to a sensitivity spectral density of $\Psi_{\rm shot}\simeq3.4\cdot10^{-9}  {\frac{1}{\sqrt{{\text Hz}}}}$. It is interesting to note that such a limit depends exclusively on the laser power before the analyzer and the quantum efficiency of the detector.

Other intrinsic noise sources are photodiode dark current noise $i_{\rm dark}=V_{\rm dark}\sqrt{\Delta\nu}/G$, Johnson current noise $i_{\rm J}=\sqrt{4K_{B}T\Delta\nu/G}$ of the transimpedence $G$ in the amplifier of the photodiode, and residual laser intensity current noise $i_{\rm RIN}=I_{0}q\cdot{\rm RIN}(\omega)\sqrt{\Delta\nu}$.
These noises must be kept below $i_{\rm shot}$ at a frequency near $\omega_{\rm Mod}$ in order to reach the theoretical sensitivity. The expressions for these noise contributions to the ellipticity spectral noise density can be obtained from equation (\ref{sensit}):
\begin{eqnarray}
\Psi_{\rm shot}&=&\sqrt{\frac{e}{I_{0}q}\left(\frac{\sigma^{2}+\zeta_{0}^{2}/2}{\zeta_{0}^{2}}\right)}\\
\Psi_{\rm dark}&=&\frac{V_{\rm dark}}{G\sqrt{2}}\frac{1}{I_{0}q\zeta_{0}}\\
\Psi_{\rm J}&=&\sqrt{\frac{2K_{B}T}{G}}\frac{1}{I_{0}q\zeta_{0}}\\
\Psi_{\rm RIN}&=&\frac{\rm RIN(\omega_{\rm Mod})}{\sqrt{2}}\frac{\sqrt{\left(\sigma^{2}+\zeta_{0}^{2}/2\right)^{2}+\left(\zeta_{0}^{2}/2\right)^{2}}}{\zeta_{0}}
\end{eqnarray}

\begin{figure}
\begin{center}
{\includegraphics*[width=8cm]{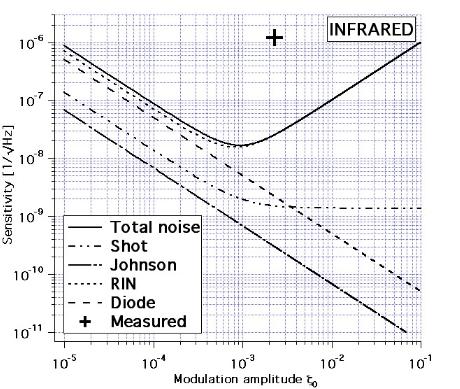}}
{\includegraphics*[width=8cm]{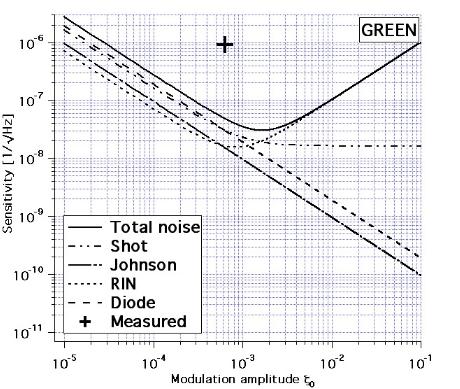}}
\caption{\it Calculated noise contributions for the infrared and green configurations of the PVLAS apparatus. See text and table \ref{parameters} for the parameters of each configuration.}
\label{noise}
\end{center}
\end{figure}

With the PVLAS experimental parameters given in Table \ref{parameters}, the contribution of each of these noises to the sensitivity spectral density can be plotted as a function of the modulation amplitude $\zeta_0$. This allows the optimization of the modulation amplitude.
Figure \ref{noise} shows the corresponding plots for the infrared and green configurations. In each graph a cross marks the current experimental sensitivity.
\begin{table}[h!]
\caption{Experimental parameters for the two laser configurations.}
\begin{tabular}{c|c|c|c|c|c|c}
\hline\noalign{\smallskip}
Configuration & Photodiode & Cavity output & Extinction & RIN($\omega_{Mod}$)  & Gain $G$ & Photodiode noise \\
 & efficiency $q$ [A/W] & power $I_{0}$ [mW] & ratio $\sigma^{2}$ & [1/$\sqrt{\rm Hz}$] & [V/A] & $V_{\rm dark}$ [$\mu$V/$\sqrt{\rm Hz}$] \\
\noalign{\smallskip}\hline\noalign{\smallskip}
Green & 0.2 & 1.5 & $5\cdot10^{-7}$ & $2\cdot10^{-5}$ & $10^{9}$ & $2$ \\
infrared & 0.7 & 60 & $5\cdot10^{-7}$ & $2\cdot10^{-5}$ & $10^{7}$ & $8$ \\
\noalign{\smallskip}\hline
\end{tabular}
\label{parameters}
\end{table}

As can be seen, in both configurations we are still well away from the theoretical limit. 
The one noise source which cannot, at the moment, be controlled is the low frequency spurious ellipticity $\alpha(t)$ (see equation (\ref{spurious})) induced in the system. We believe this noise is due to the movement of the granite tower. Since the mirrors and optical elements have a structural birefringence `map' with a gradient \cite{Micossi}, these movements will generate ellipticity noise. Indeed, we have measured the induced ellipticity as a function of movement at the top of the tower and found a value $\simeq 0.4$ m$^{-1}$. To reach the theoretical limit, the relative movement of the top of the tower respect to the lower optical bench must be less than $10^{-8}$~m/$\sqrt{\rm Hz}$.
\section{Previous photon-photon scattering results}
In this section we review the results obtained by D. Bernard {\em et al.} in a direct search for diffused photons from two colliding beams \cite{Bernard} and the results from the BFRT collaboration \cite{Cameron}, another experiment designed to search for axions via ellipsometric techniques. There are several proposals for improving the direct scattering measurements \cite{Brodin,exawatt,Luiten1,Luiten2} and for detecting the magnetic birefringence of vacuum \cite{bmv,ni,pugnat}.

\subsection{Stimulated photon-photon scattering}
In \cite{Bernard} direct photon-photon scattering was searched for. Differently from what was previously done \cite{BernardOld}, the authors searched for stimulated scattering when three high power ultra short beams were crossed. The advantage of this configuration is to fix the angle and wavelength for the scattered photon: scattered photons are searched for in a defined direction and at a defined wavelength.
Indeed in two-beam scattering, energy (indicated as $e_i$) and 3-momentum (indicated as ${\bf k}_{\rm i}$) must be conserved: $e_{\rm 1} + e_{\rm 2} = e_{\rm 3} + e_{\rm 4}$; {\bf k}$_{\rm1}$ + {\bf k}$_{\rm2}$ = {\bf k}$_{\rm3}$ + {\bf k}$_{\rm4}$.
This leaves two free parameters for the final state. In the three beam configuration the fourth (scattered) beam must satisfy the condition  {\bf k$_{\rm4}$}={\bf k$_{\rm1}$} + {\bf k$_{\rm2}$} - {\bf k$_{\rm3}$} and $\lambda_{\rm4} = \left(1/\lambda_{\rm1} + 1/\lambda_{\rm2} - 1/\lambda_{\rm3}\right)^{-1}$. In the configuration being discussed $\lambda_{\rm1} = \lambda_{\rm2} = 800$ nm and $\lambda_{\rm3} = 1300$~nm, resulting in $\lambda_{\rm4} = 577$~nm.

The theoretical analysis of three wave mixing in vacuum develops in analogy with three wave mixing in a medium \cite{Bernard}. In this latter case the medium polarizability is written as
\begin{equation}
{\cal{P}}(t) = \underline{\chi}^{(1)}{\cal E}(t) + \underline{\underline{\chi}}^{(2)}{\cal E}^{2}(t) + \underline{\underline{\underline{\chi}}}^{(3)}{\cal E}^{3}(t)+\cdots
\end{equation}

In four wave mixing the authors show that the growth rate of the electric field ${\cal E}_{04}$ of the scattered beam depends on $\chi^{(3)}$ and can be written as 
\begin{equation}
\frac{d{\cal E}_{04}}{dz}=-\frac{i\omega_{4}}{2c}\chi^{(3)}{\cal E}_{01}{\cal E}_{02}{\cal E}_{03} \qquad\mbox{with}\qquad \frac{d}{dz}=\frac{\partial}{\partial z}+\frac{1}{c}\frac{\partial}{\partial t}.
\label{growth}
\end{equation}

When considering the Euler-Heisenberg Lagrangian density correction one finds that the growth rate of ${\cal E}_{04}$ in vacuum has the same form as equation (\ref{growth}) with the QED susceptibility of vacuum having the expression
\begin{equation}
{\chi^{(3)}}={\chi_{v}}^{(3)}=\frac{2\hbar e^{4}K}{360\pi^{2}m^{4}c^{7}\epsilon_{0}}=A_{e}\frac{2K}{c^2}\simeq3.0\cdot10^{-41}K\mbox{~m}^2/\mbox{V}^2
\end{equation}
where $K$ is a parameter depending on the direction of the incident beams and their polarization. In the experiment reported in \cite{Brodin} $K\simeq0.56$.

Integration of equation (\ref{growth}) leads to a number of counts per pulse crossing
\begin{equation}
N_{4}=\epsilon_{PM}\epsilon_{Sp}\epsilon_{loss}\frac{128}{\pi\sqrt{3}^{3}}\frac{(\hbar\omega_{4})E_{1}E_{2}E_{3}}{e^{4}w^{2}(c\tau)^{2}}(\chi^{(3)})^{2}
\end{equation}
where $E_{i}$ are the energies of the three incoming laser pulses, $\epsilon_{PM}$,$\epsilon_{Sp}$, $\epsilon_{loss}$ are the quantum efficiency of the photomultiplier tube, the 
transmission of the spectrometer and a loss factor due to a beam position oscillation, and $w$ and $c\tau$ are respectively the beam waist and bunch length.
The value of the third order susceptibility $\chi^{(3)}$ of nitrogen was measured and compared to other experiments. An order of magnitude agreement was observed, allowing a calibration of the apparatus. A comparison between the expected vacuum counts calculated from QED and the observed counts resulted in a limit on the total photon-photon cross section at 0.8~eV center of mass energy of 
\begin{equation}
\sigma_{\gamma\gamma}^{\rm (Bernard)}=\frac{N_{4,{\rm obs}}}{N_{4,{\rm QED}}}\sigma_{\gamma\gamma}^{\rm (QED)}=1.5\cdot10^{-52} \text{ m}^2
\end{equation}
which is eighteen orders of magnitude larger than the theoretical QED cross section.
\subsection{Brookhaven-Fermilab-Rochester-Trieste (BFRT) results}
The principle of the ellipsometer in the BFRT collaboration is the same as the one shown in Figure \ref{SchemePVLAS}. In this case the cavity was a multi-pass cavity with a number of reflections which varied from 34 to 578. The laser wavelength was $514.5$~nm and the length of the magnetic field region was $8.8$~m. To modulate the magnetic vacuum birefringence the magnetic field was ramped from $2.63$ to $3.87$~T at a frequency of $30$~mHz. The sensitivity of the apparatus varied as a function of the number of reflections in the multi-pass cavity and consequently did the final limit on the acquired ellipticity. The results are summarized in table \ref{ResultsBFRT}.

\begin{table}[h!]
\caption{Summary of the BFRT experimental parameters and results together with the limit achieved on the parameter $A_{e}$.}
\begin{tabular}{c|c|c|c}
\hline\noalign{\smallskip}
Number of & Measured & Ellipticity upper & $A_e$ Upper \\
passes & sensitivity [$1/\sqrt{{\text Hz}}$]& bound  $\psi_{\rm limit}$ at 95\%C.L. & bound [$T^{-2}$]\\
\noalign{\smallskip}\hline\noalign{\smallskip}
0 (shunt) & $2.6\cdot10^{-8}$& $7.7\cdot10^{-10}$ & n.a.\\
$34$& $7.9\cdot10^{-8}$ & $2.0\cdot10^{-9}$ & $1.4\cdot10^{-19}$\\
$578$ & $1.5\cdot10^{-6}$ & $5.1\cdot10^{-8}$ & $2.1\cdot10^{-19}$\\
\noalign{\smallskip}\hline
\end{tabular}
\label{ResultsBFRT}
\end{table}
Due to the fact that the magnetic field was ramped around a central value $B_{0} = 3.25$~T, with an excursion $\pm\Delta B = \pm0.62$~T, the expression for the Euler-Heisenberg induced magnetic birefringence is
\begin{equation}
\Delta n = 3A_{e}(2B_{0}\Delta B)
\end{equation}
and the limit on $A_e$ attainable from the BFRT results is 
\begin{equation}
A_e<\psi_{\rm limit}\frac{\lambda}{6B_{0}\Delta B N L}
\end{equation}

When translated into a photon-photon cross section for unpolarized light these results give a limit of
\begin{equation}
\sigma_{\gamma\gamma}^{(\rm BFRT)}<1.6\cdot10^{-57} \text{ m}^2
\end{equation}
This must be compared to the QED photon-photon cross section at the same wavelength of 514.5~nm which is $\sigma_{\gamma\gamma}^{\rm (QED)}=1.44\cdot10^{-67}$~$\text{ m}^2$.

\section{PVLAS Results}
Gas measurements, for calibration, and vacuum birefringence measurements were conducted with the apparatus in both the infrared and green configurations. We present here measurements taken with the magnet energized at 2.3~T. This choice of field strength is motivated by the strong suppression of the stray field outside the magnet. Indeed, at higher fields the presence of a stray field has resulted in a yet to be understood spurious ellipticity signal \cite{PRD}. The total integration times were $T_{1064} = 45200$~s at 1064~nm and $T_{532} = 28300$~s at 532~nm. 

\subsection{Gas calibration measurements}
\begin{figure}[htb]
\begin{center}
{\includegraphics*[width=8cm]{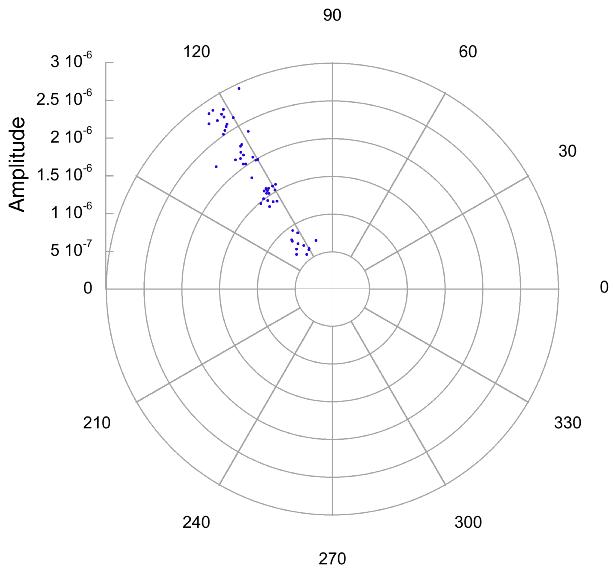}}
\caption{\it Polar plot for the ellipticity signal generated with a 2.3~T magnetic field intensity when Helium gas is present in the vacuum chamber. The figure shows the signal for four different gas pressures: 5, 10, 15 and 20 mbar. Each data point represents amplitude and phase of the signal peak observed in a 100 s long time record.}
\label{CM}
\end{center}
\vspace{-0.6cm}
\end{figure}

Calibration of the ellipsometer is done by taking advantage of the Cotton-Mouton effect \cite{rizzo} in gases. In the presence of an external magnetic field perpendicular to the propagation of a light beam, gases become birefringent. Depending on the gas, the induced birefringence may be positive ($n_{\parallel}-n_{\perp} > 0$; e.g. He) or negative ($n_{\parallel}-n_{\perp} < 0$; e.g. N$_2$). These measurements also allow the verification of the Fourier phase of the sidebands of $\omega_{\rm Mod}$ at $\omega_{\rm Mod}\pm2\Omega_{\rm Mag}$ with what they should be (see table \ref{components}). Indeed, the ellipticity induced by a birefringence is maximum when the angle between the polarization and the slow axis defined by the magnetic field is $45^{\circ}$ ($n_{\parallel}-n_{\perp} > 0$). Figure \ref{CM} shows a polar plot corresponding to the amplitude and phase of the signal demodulated at $\omega_{\rm Mod}$ due to Helium gas at four different pressures (5, 10, 15 and 20~mbar), measured with a field intensity of 2.3~T. As can be seen, the experimental values lie on a straight line with a Fourier phase of $125^{\circ}$. This defines the physical axis of signals. A gas with a negative Cotton-Mouton constant would generate a signal at $180^{\circ}$ with respect to the signals shown in Figure \ref{CM}.

Different gases were measured \cite{O2,Xe} resulting in an accuracy better than 20\%.

\subsection{Vacuum measurements}
\begin{figure}[htb]
\begin{center}
{\includegraphics*[width=7.5cm]{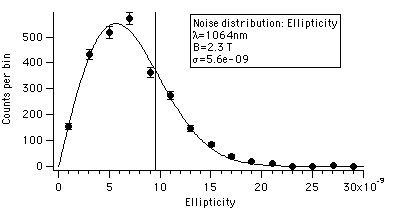}
\includegraphics*[width=7.5cm]{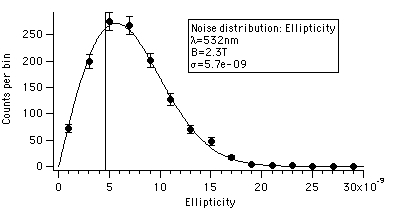}}
\caption{\it Noise distributions in the magnet rotation frequency band $1.92~\Omega_{Mag}$~-~$2.08~\Omega_{Mag}$ for the 2.3~T ellipticity measurements with the IR (left) laser and the green (right) laser. The vertical line indicates the value in the Fourier spectrum corresponding to $2\Omega_{Mag}$. Indicated in the legend is the value of $\sigma$ for the two wavelengths.}
\label{rayleighBirlow}
\end{center}
\vspace{-0.6cm}
\end{figure}

The complete data sets of the signals from the lock-in amplifiers demodulated at $\omega_{\rm Mod}$ were analysed by a Fourier transform. No peak was found at $2\Omega_{\rm Mag}$ as would be expected from a magnetically induced ellipticity. The data will therefore be presented as a noise histogram in a frequency band around $2\Omega_{\rm Mag}$, between $1.92~\Omega_{\rm Mag}$ and $2.08~\Omega_{\rm Mag}$ (see Figure~\ref{rayleighBirlow}).

The probability density function for the Fourier amplitude $r_{\rm F}=\sqrt{x_{\rm F}^{2}+y_{\rm F}^{2}}$, where $x_{\rm F}$ and $y_{\rm F}$ are the projections of the Fourier transform along the physical and the non physical phases (defined above) respectively,  is given by the Rayleigh distribution, $p(r_{\rm F}) = r_{\rm F}e^{-\frac{r_{\rm F}^2}{2\sigma^2}}/\sigma^2$. In this expression $\sigma$ is the standard deviation of the Gaussian noise distributions of $x_{\rm F}$ and $y_{\rm F}$ from which we deduced our limits on the induced ellipticity. To extract $\sigma$ from the histograms the $r_{\rm F}$ noise distributions were fitted with the Rayleigh distribution.

The 95\% confidence limits are then deduced from the cumulative distribution $P(r_{\rm F}) = 1-e^{-r_{\rm F}^2/2\sigma^2}$.
In Figure \ref{rayleighBirlow} are shown the histograms and fits for the measurements taken with the 1064~nm and 532~nm lasers. Superimposed on these graphs, represented by a vertical black line, are the values at the bins corresponding to $2\Omega_{Mag}$ of the Fourier spectrum of the demodulated signal. As can be seen, these are well within the noise distributions. The standard deviations of the two distributions are very similar even though the integration time with the 532~nm laser was 70\% of the integration time with the 1064~nm. This is due to the lower noise encountered with the 532~nm setup.

Table \ref{tab1} gives the $95\%$ confidence level background values for the ellipticity measurements with the PVLAS and the BFRT apparatus. The stimulated scattering results are also reported where appropriate. The parameters for the different configurations are also reported in the same table. In the last three columns we also report the limits on $A_e$, $\sigma_{\gamma\gamma}$ and the ratio ${\sigma_{\gamma\gamma}}/{\sigma_{\gamma\gamma}^{\rm (QED)}}$. \begin{table}
\begin{tabular}{c|c|c|c|c|c|c}
\hline\hline
Measurement & Photon & noise & $(B^{2}l)_{\rm equiv}$ & $A_e$ bounds & $\sigma_{\gamma\gamma}$ bounds & \\
type & energy & floor  & [T$^2$m]  & 95\% C.L. [T$^{-2}$] &[m$^{2}$] & ${\sigma_{\gamma\gamma}}/{\sigma_{\gamma\gamma}^{\rm (QED)}}$ \\\hline
stimulated scatter \cite{Bernard} & $0.8$~eV c.o.m & & &$1.2\cdot10^{-15}$ &  $1.5\cdot10^{-52}$ & $8\cdot10^{17}$\\
BFRT Ellipticity \cite{Cameron} & $2.42$~eV & $4.9\cdot 10^{-9}$  & 1197 & $1.4\cdot10^{-19}$ &  $1.6\cdot10^{-57}$ & $11\cdot10^{9}$\\
PVLAS Ellipticity \cite{PRD} & $1.17$~eV & $1.4\cdot 10^{-8}$ & 238000 & $6.6\cdot10^{-21}$ & $4.6\cdot10^{-62}$& $2.5\cdot10^{7}$\\
PVLAS Ellipticity & $2.34 $~eV & $1.4\cdot 10^{-8}$ & 124000 & $6.3\cdot10^{-21}$ & $2.7\cdot10^{-60}$ & $2.3\cdot10^{7}$\\

\hline\hline
\end{tabular}
\caption{Ellipticity results and stimulated scatter results for $\sigma_{\gamma\gamma}$}
\label{tab1}
\end{table}

Although experimental results have not reached the predicted QED values, bounds have been improved. We believe that at the moment the sensitivity is limited by seismically induced spurious ellipticities.

\section{Discussion and conclusions}
We have reported here the interpretation of vacuum magnetic birefringence limits in terms of photon-photon scattering. Although the sensitivity of our apparatus has not reached its theoretical shot noise limit, the ellipsometric technique is at the moment the most sensitive one for approaching low energy non linear QED effects. In a general post-Maxwellian framework, though, direct scattering measurements are necessary to extract the free parameters $\xi\eta_{1}$ and $\xi\eta_{2}$ (see equation (\ref{Lpm})). In the Euler-Heisenberg framework we are now a factor 4800 away from the theoretical parameter, $A_{e}=1.32\cdot10^{-24}$~T$^{-2}$, describing non linear quantum electrodynamic effects: 
\begin{equation}
A_{e}^{\text{(Exp.)}}<6.3\cdot10^{-21} {~\text T}^{-2} {\text{~@~95\% C.L.}}
\end{equation}

Always in the Euler-Heisenberg framework, from the experimental bound on $A_{e}$ one can place the following upper bounds on the photon-photon cross section for non polarized light in the limit $\hbar\omega \ll m_{e}c^{2}$, at 1064~nm and 532~nm respectively of:
\begin{eqnarray}
\sigma_{\gamma\gamma}^{(1064)} < 4.6\cdot10^{-62} {\text {~m}}^2\\
\sigma_{\gamma\gamma}^{(532)} < 2.7\cdot10^{-60} {\text {~m}}^2
\end{eqnarray}

\end{document}